\documentclass[prl,aps,twocolumn,superscriptaddress,showpacs]{revtex4-1}

\usepackage{graphicx}
\usepackage{epstopdf}

\begin{document}

\title{From soft harmonic phonons to fast relaxational dynamics in CH$_{3}$NH$_{3}$PbBr$_{3}$}

\author{I.P. Swainson}
\affiliation{National Research Council, Chalk River, Ontario, KOJ 1JO, Canada}

\author{C. Stock}
\affiliation{School of Physics and Astronomy, University of Edinburgh, Edinburgh EH9 3JZ, UK}

\author{S. F. Parker}
\affiliation{ISIS Facility, Rutherford Appleton Laboratory, Chilton, Didcot OX11 0QX, UK}

\author{L. Van Eijck}
\affiliation{Institut Laue-Langevin, 6 rue Jules Horowitz, Boite Postale 156, 38042 Grenoble Cedex 9, France}
\affiliation{Reactor Institute Delft, Delft University of Technology, 2629JB Delft, Netherlands}

\author{M. Russina}
\affiliation{Helmholtz Zentrum Berlin fur Materialien und Energie GmbH, D-14109 Berlin, Germany}

\author{J. W. Taylor}
\affiliation{ISIS Facility, Rutherford Appleton Laboratory, Chilton, Didcot OX11 0QX, UK}
\affiliation{European Spallation Source, Universitetparken 1, 2100, Copenhagen, Denmark}

\date{\today}

\begin{abstract} 

The lead-halide perovskites, including CH$_{3}$NH$_{3}$PbBr$_{3}$, are components in cost effective, highly efficient photovoltaics, where the interactions of the molecular cations with the inorganic framework are suggested to influence the electronic and ferroelectric properties. CH$_{3}$NH$_{3}$PbBr$_{3}$ undergoes a series of structural transitions associated with orientational order of the CH$_{3}$NH$_{3}$ (MA) molecular cation and tilting of the PbBr$_{3}$ host framework.  We apply high-resolution neutron scattering to study the soft harmonic phonons associated with these transitions, and find a strong coupling between the PbBr$_{3}$ framework and the quasistatic CH$_{3}$NH$_{3}$ dynamics at low energy transfers.  At higher energy transfers, we observe a PbBr$_{6}$ octahedra soft mode driving a transition at 150 K from bound molecular excitations at low temperatures to relatively fast relaxational excitations that extend up to $\sim$ 50-100 meV.  We suggest that these temporally overdamped dynamics enables possible indirect band gap processes in these materials that are related to the enhanced photovoltaic properties.

\end{abstract}


\maketitle

Organic-inorganic, hybrid perovskites (OIPs) are materials based upon an inorganic perovskite host framework with an organic molecular cation occupying the interstitial space.   These materials have been studied for quite some time~\cite{Wyckoff28:1}, but interest has recently surged owing to their use in photovoltaic devices and to possible ferroelectricity.~\cite{Kojima09:131} While earlier work centered on Sn-based OIPs as possible sensor materials~\cite{Mitzi01:45,Mitzi07:48}, recent interest has been focused towards Pb-based OIPs, due to their potential advantages in inexpensive photovoltaic devices, with efficiencies of the order of 20\%.~\cite{Park13:4}  We apply neutron scattering to study the coupled dynamics of the host PbBr$_{3}$ framework and the methylamonium (MA) cation in CH$_{3}$NH$_{3}$PbBr$_{3}$, and discuss the possible relation with the photovoltaic properties.  We map out the soft phonons and low-temperature quasistatic molecular rotations, and show that the low-temperature, harmonic fluctuations cross over to temporally overdamped dynamics at high temperature.

\begin{figure}[t]
\includegraphics[width=8.3cm] {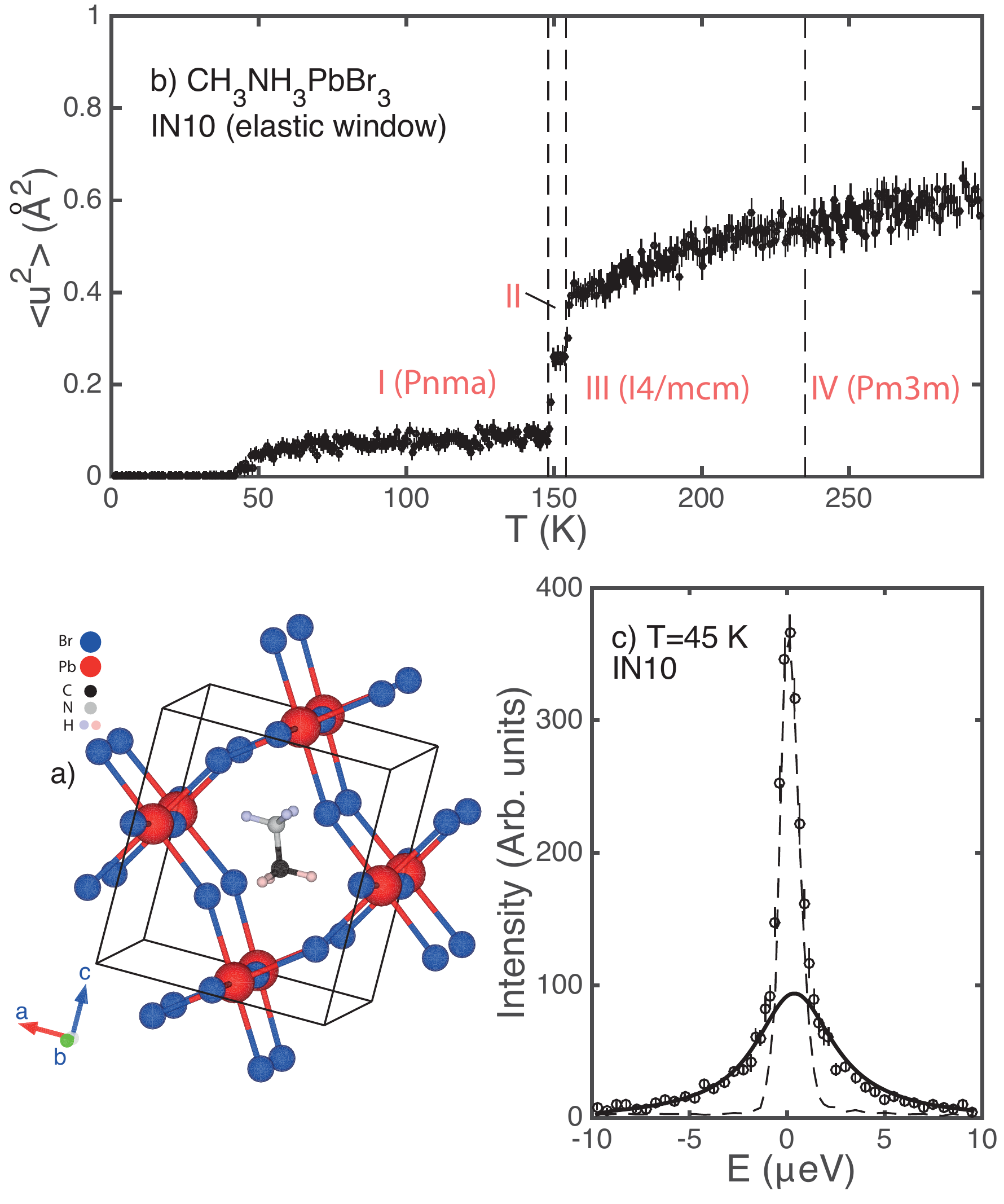}
\caption{\label{figure1}  $(a)$ The structure of CH$_{3}$NH$_{3}$PbBr$_{3}$ illustrating the MA molecular cations in the PbBr$_{3}$ framework at 5 K.  $(b)$ shows the mean-squared displacement of the hydrogen atoms ($\langle u^{2} \rangle$) extracted from an IN10 elastic window scan as a function of temperature and $(c)$ plots a representative quasielastic scan taken on IN10 at 45 K.}
\end{figure}

The OIPs are composed of two sublattices (Fig. \ref{figure1} (a) for CH$_{3}$NH$_{3}$PbBr$_{3}$): the inorganic sublattice, consisting of a fully corner-bonded framework of octahedra (PbBr$_{6}^-$); and the organic sublattice consisting of the MA molecular cation, (CH$_{3}$NH$_{3}^+$).  At high temperatures, the structure of CH$_{3}$NH$_{3}$PbBr$_{3}$ is cubic (space group Pm3m), and below 235~K has a tetragonal structure in symmetry I4/mcm.~\cite{Swainson03:176,Ong03:176}  At 150~K, a transition to an unknown structure (believed to be incommensurate~\cite{Futterer95:7}) occurs followed by further distortion to orthorhombic (Pnma) at 148~K.~\cite{Knop90:68,Worhatch08:20,Mashiyama02:71,Mashiyama07:348} Neutron diffraction measurements have suggested that these transitions originate from tilting of the PbBr$_{6}$ octahedra and orientational ordering of the MA cation.~\cite{Swainson03:176,Chi05:178}  We note that despite minor differences in the phase diagrams, for all MAPbX$_{3}$ (X=Cl,Br,I), a transition to an ordered phase, in which the octahedra are tilted and the MA cations have relatively well-defined orientations, occurs in the temperature range 150--175~K~\cite{Knop90:68,Swainson03:176,Chi05:178,Mashiyama07:51}.  

The MA orientations have  been implicated as influencing charge separation through a bulk polarization.~\cite{Quarti14:26}  This, in turn, has been suggested to enhance the the photovoltaic properties.~\cite{Zheng15:6} To this end, we apply neutron inelastic scattering to study the dynamics of the MA molecular cation, the PbBr$_{3}$ framework, and the coupling between the two sets of dynamics.  The experiments were performed on the MARI (ISIS, UK), NEAT (HMI, Germany), IN4,  and IN10 (ILL, France) spectrometers.   Samples of CH$_{3}$NH$_{3}$PbBr$_{3}$ and CD$_{3}$ND$_{3}$PbBr$_{3}$ were synthesized as discussed elsewhere.~\cite{Swainson03:176}  The scattering measurements of the harmonic vibrations were compared against Castep/Aclimax first-principles calculations.~\cite{CASTEP,ACLIMAX}

The low-temperature phase of CH$_{3}$NH$_{3}$PbBr$_{3}$ consists of a PbBr$_{3}$ framework with a MA molecular cation hydrogen bonded to the Br atoms (Fig. \ref{figure1} $(a)$).~\cite{Swainson03:176}  Because the neutron incoherent cross section of H is over an order of magnitude larger than those of the non-H atoms, the angular-independent neutron scattering intensity is dominated by H dynamics.   We exploit this in Fig. \ref{figure1} $(b)$ which plots the average mean-squared displacement of the hydrogen atoms.  This is extracted from an energy integrating ``fixed window" scan integrating over energy transfers of $\pm$ 1 $\mu eV$ (further details in the supplementary information~\cite{supp}).  The temperature scans illustrate an onset of fluctuations near 50 K where dynamic processes enters the resolution of the IN10 spectrometer.  With increasing temperature, this is followed by an abrupt change at the I--II and II-III structural transitions of 148 K and 154 K, respectively, from where the mean-squared displacement evolves continuously with increasing temperature.

The discontinuity in $\langle u^{2} \rangle$ at 148 K (phase I--II) has been previously noted with neutron diffraction~\cite{Swainson03:176}, calorimetry, and NMR.~\cite{Knop90:68}  Based on comparisons to MAPbCl$_3$ and TMAGeCl$_3$ (TMA=tetramethylammonium), it is likely that this is a transition to an intermediate incommensurate phase separating orthorhombic (phase I) and tetragonal (phase III) states.~\cite{Futterer95:7,Kawamura99:35,Swainson05:61}   There is no observable discontinuity in the mean-squared displacement of the hydrogen atoms associated with the III--IV (tetragonal to cubic) transition on this timescale ($1/ \tau \sim \delta E=\pm$ 1 $\mu eV$). This indicates the low-energy molecular dynamics are similar over this temperature range.

The timescale of the dynamics of the MA ion can be probed from energy analyzing neutron quasielastic scattering (QENS). Figure \ref{figure1} $(c)$ shows a representative QENS scan taken on IN10 at 45 K displaying a two-component line shape with a sharp, resolution-limited elastic component (dashed line) and a broader-than-resolution relaxational component, fit by a lorentzian (solid curve).  The broader lorentzian is indicative of molecular dynamics from which a timescale ($\tau$) can be extracted from the energy width ($\gamma \sim 1 / \tau$).   We now discuss the quasielastic components as a function of both momentum and temperature.

\begin{figure}[t]
\includegraphics[width=8.2cm] {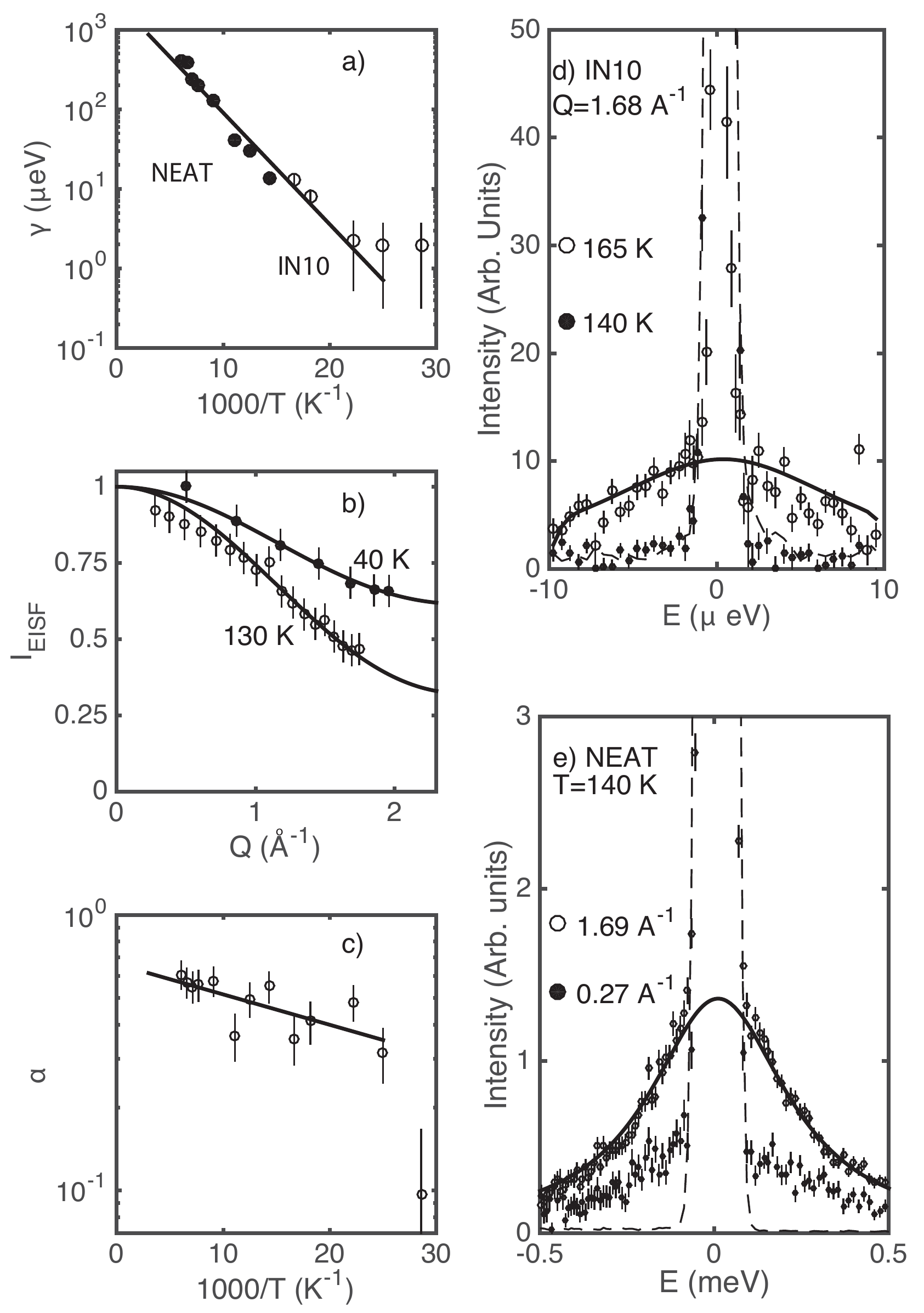}
\caption{\label{QENS_summary}  The quasielastic linewidth $(a)$, momentum dependence $(b)$, and the elastic fraction $(c)$ are plotted for CH$_{3}$NH$_{3}$PbBr$_{3}$.  Sample quasielastic scans and fits are shown from the IN10 backscattering spectrometer $(d)$ and the NEAT direct-geometry time-of-flight spectrometer $(e)$.}
\end{figure}
x \ref{QENS_summary} plots the energy ($E$), momentum ($Q$), and temperature ($T$) dependence of the quasielastic scattering in the CH$_{3}$NH$_{3}$PbBr$_{3}$.  $T$- and $Q$-dependent parameters were extracted by fitting each energy scan to the sum of an elastic and dynamic component ($S(Q,E)=I_{\rm el}+I_{\rm qe}=I_{\rm elastic}\delta(E)+I_{\rm dynamic}/(1+(E/\gamma)^{2})$) convolved with the resolution.  Following previous studies, we define the elastic incoherent structure factor (EISF) from the normalized intensity $I_{\rm EISF}=I_{\rm elastic}/(I_{\rm elastic}+I_{\rm dynamic})$.~\cite{Line94:21}  Example fits are displayed in Fig. \ref{figure1} $(c)$ and Fig. \ref{QENS_summary} $(d,e)$.

The $Q$ dependence of the EISF is shown in Fig. \ref{QENS_summary} $(b)$, and decays monotonically with $Q$ at all temperatures studied (shown in Fig. \ref{QENS_summary} $(e)$).  As noted in Ref. \onlinecite{Bee:book}, the $Q$ dependence of the EISF is sensitive to the real space nature of the equilibrium molecular motions; however, distinguishing features in the models are only present at momentum transfers beyond the range accessible in this experiment limited by strong nuclear Bragg peaks,  resolution, and kinematics of neutron spectroscopy.   We have therefore followed Ref. \onlinecite{Line94:21} and fit the results to a model for isotropic molecular reorientations with the temperature dependence given by two components - one where the molecular cation is bound, and a second fraction ($\alpha$) that is fluctuating and contributes to the inelastic scattering component.  The cross section, therefore, has the form $I_{\rm total}=(1-\alpha)I_{\rm bound}+\alpha(I_{\rm el}+I_{\rm qe})$.   ``Isotropic" molecular reorientations give a dependence of $I_{\rm EISF}(Q)=\sin(Qr)/Qr$, with $r$ the radius of the molecular cation and for simplicity we have taken $r$ to be the C--H distance of 1.05 \AA\ from the methyl group.  We do not consider whole body rotations which would require a larger effective radius including the C--N distance~\cite{Kieslich14:5} which is inconsistent with the momentum dependence.  The bound component (I$_{\rm bound}$) is taken to be a constant independent of momentum transfer.  Therefore, we consider rigid motions around a fixed equilibrium position where the molecular cations are orientationally ordered as expected based on previous structural analysis.~\cite{Swainson03:176}  The fraction of MA ions in motion is given by $\alpha$ in Fig. \ref{QENS_summary} $(c)$.   A fit to an activation form $\alpha \propto e^{-E_{\rm B}/T}$ gives $E_{\rm B}$=51 $\pm$ 5~K which characterizes the energy required to activate molecular reorientation which matches the onset temperature of observable displacements of the hydrogen atoms in Fig. \ref{figure1} $a)$.  The T-dependence of $\alpha$ (Fig. \ref{QENS_summary} $c$) implies a phase transition at $\sim$ 150 K when approximately 50 \% of the molecules become ``unbound". 

The timescale of the molecular reorientations is presented in Fig. \ref{QENS_summary} $(a)$ plotting  $\gamma$ vs T below 150 K.  The molecular motions above 150 K were too fast for the dynamic ranges on both spectrometers and are discussed further below.   The solid curve is a fit to an Arrhenius form $\gamma=\gamma_{0} e^{-E_{\circ}/T}$ with $\gamma_{0}$=323 $\pm$ 20 K and $E_{\circ}$=2.3 $\pm$ 0.1 meV.  The deviation at low temperatures is due to the resolution limit of $\sim$ 2 $\mu eV$ on IN10 from which all data appear resolution limited within error as the dynamics leave the spectrometer window.   Our analysis reveals two energy scales, one lower energy E$_{\rm B}$ characterizing the energy scale to allow molecular motion and a second higher activation energy $\gamma_{0}$ associated with the energy scale of the molecular reorientation.

An unusual feature of the data, shown in Fig. \ref{QENS_summary} $(d)$, is the slow quasistatic component which enters into the IN10 time window near the structural transition at $\sim$ 150~K.  Panel $(d)$ shows quasielastic data at 140~K and 165~K.  The 140~K data shows no inelastic contribution and is well described by the resolution function (dashed line).  The $T$=165~K shows a quasistatic contribution with a linewidth of  $\sim$ 5 $\mu eV$ with at slightly higher temperatures showing this component to be out of the time window on IN10.   Based on infrared work on OIPs~\cite{Varma92:48} and also quasielastic scattering in nitromethane ~\cite{Trevino80:73}, we expect the lower-temperature, quasistatic dynamics that is onset above $\sim$ 50~K to be dominated by slowing methyl-group rotations, whereas the dynamics observed just below the 150~K transition are likely due to the NH$_3^+$ group, which are hydrogen-bonded to the Br framework at lower temperatures. 


\begin{figure}[t]
\includegraphics[width=8.4cm] {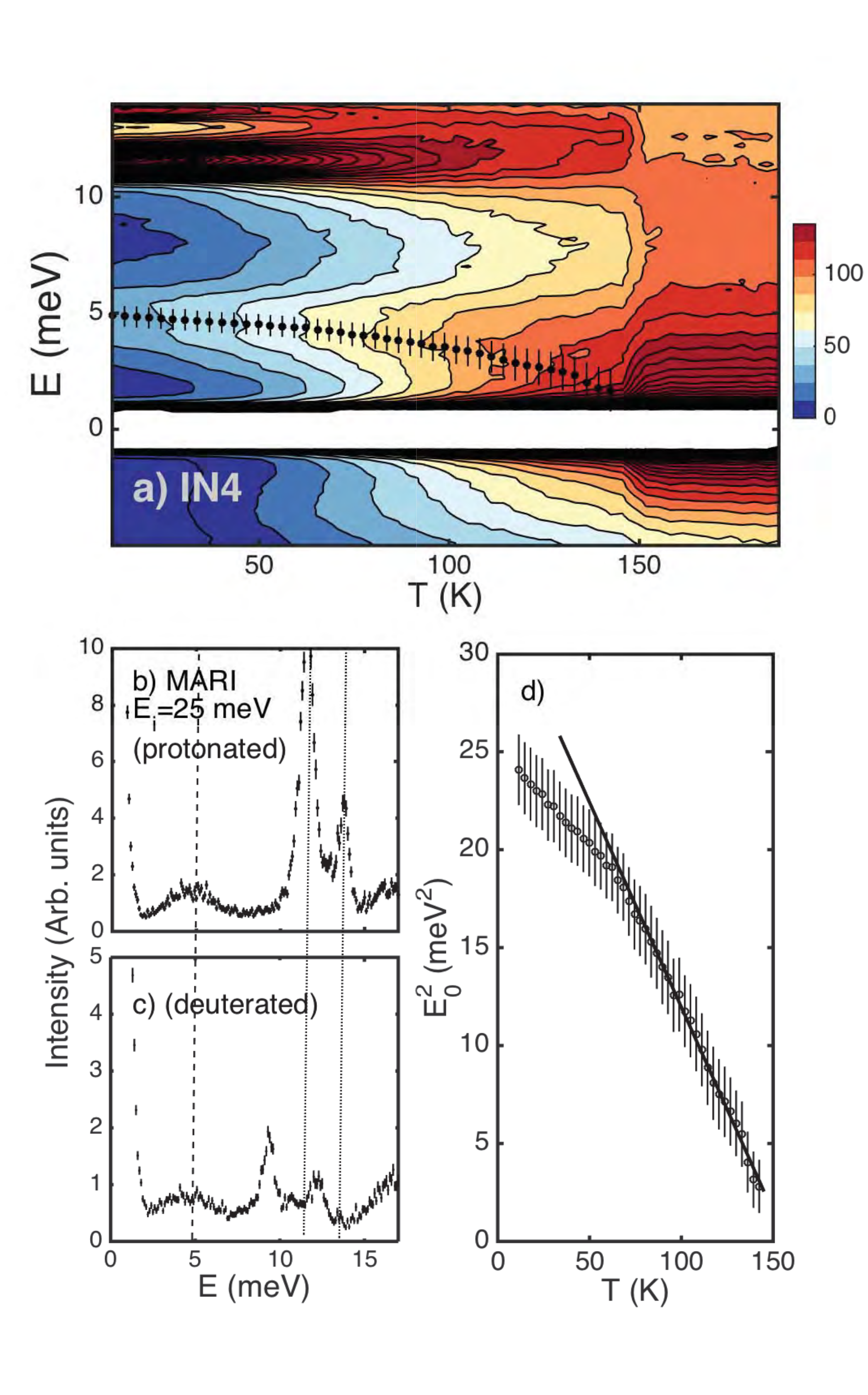}
\caption{\label{soft_mode}  $(a)$ displays inelastic scans taken on the IN4 spectrometer.  $(b)$ and $(c)$ show momentum integrated scans on MARI on fully protonated and deuterated samples. $(d)$ a plot of the square of the PbBr$_{6}$ octahedra phonon frequency as a function of $T$.}
\end{figure}

Having discussed the molecular dynamics, we now investigate the coupling between the molecular rotations and the inorganic framework.  Figure \ref{soft_mode} $(a)$ shows $Q$-integrated and powder-averaged data taken on the IN4 spectrometer mapping out the $T$ dependence of the low-energy modes.  The plot shows a series of intense modes above an energy transfer of $\sim$ 10 meV which disappear at the structural transition at 150~K.  The modes, which are sharp in energy, are replaced by a broad relaxational line shape centered at the elastic position.  This onset of fast relaxational dynamics maybe the origin of the short-range distortions of the cubic framework observed in pair distribution function studies of bulk and thin film materials.~\cite{Worhatch08:20,Choi13:14} 

Concomitant with this rapid decay, a softening of a weaker band of excitations is observed at $\sim$ 5 meV which is tracked by the solid points in Fig. \ref{soft_mode} $(a)$ obtained by fitting each spectra to a T dependent gaussian.  To assign this band to either molecular motions (MA) or the host lattice (PbBr$_{6}$ octahedra), we have performed a scan on the MARI spectrometer on both protonated and deuterated variants with the results displayed in $Q$-integrated energy scans in panels $(b)$ and $(c)$.  On deuteration, two effects are observed; first, a comparative reduction of peak intensities at $\sim$ 10 meV, due to the lack of H atoms and the corresponding smaller cross section; second, a significant shift to lower energies owing to the heavier mass of D over H.  For these two reasons, we assign the 10-meV modes to  molecular motions and the 5-meV band to PbBr$_{3}$ framework motions (not involving MA motion).    We corroborate this conclusion below by comparing MARI spectra with results from first-principles calculations.

Figure \ref{soft_mode} $(d)$ illustrates the frequency squared of the PbBr$_{6}$ octahedra mode as a function of temperature with the solid line a fit to the ``Cochran  law"~\cite{Cochran60:9} $E_{0}^{2} \propto (T-T_{\rm c})$, where $T_{\rm c}$ is the structural transition.  This follows other known zone-boundary transitions like KMnF$_{3}$~\cite{Gesi72:5} and SrTiO$_{3}$~\cite{Shirane177:69}, except that we do not observe a high-temperature recovery; instead, the spectra are dominated by the relaxational dynamics discussed above which masks any observable recovery of the soft mode.  A strong deviation from linearity is observed at $\sim$ 50~K, which is the temperature scale where molecular motions is onset (Fig. \ref{figure1}). The IN4 results illustrate a coupling between the dynamics of both molecular (organic) and framework (inorganic) sublattices.

\begin{figure}[t]
\includegraphics[width=8.5cm] {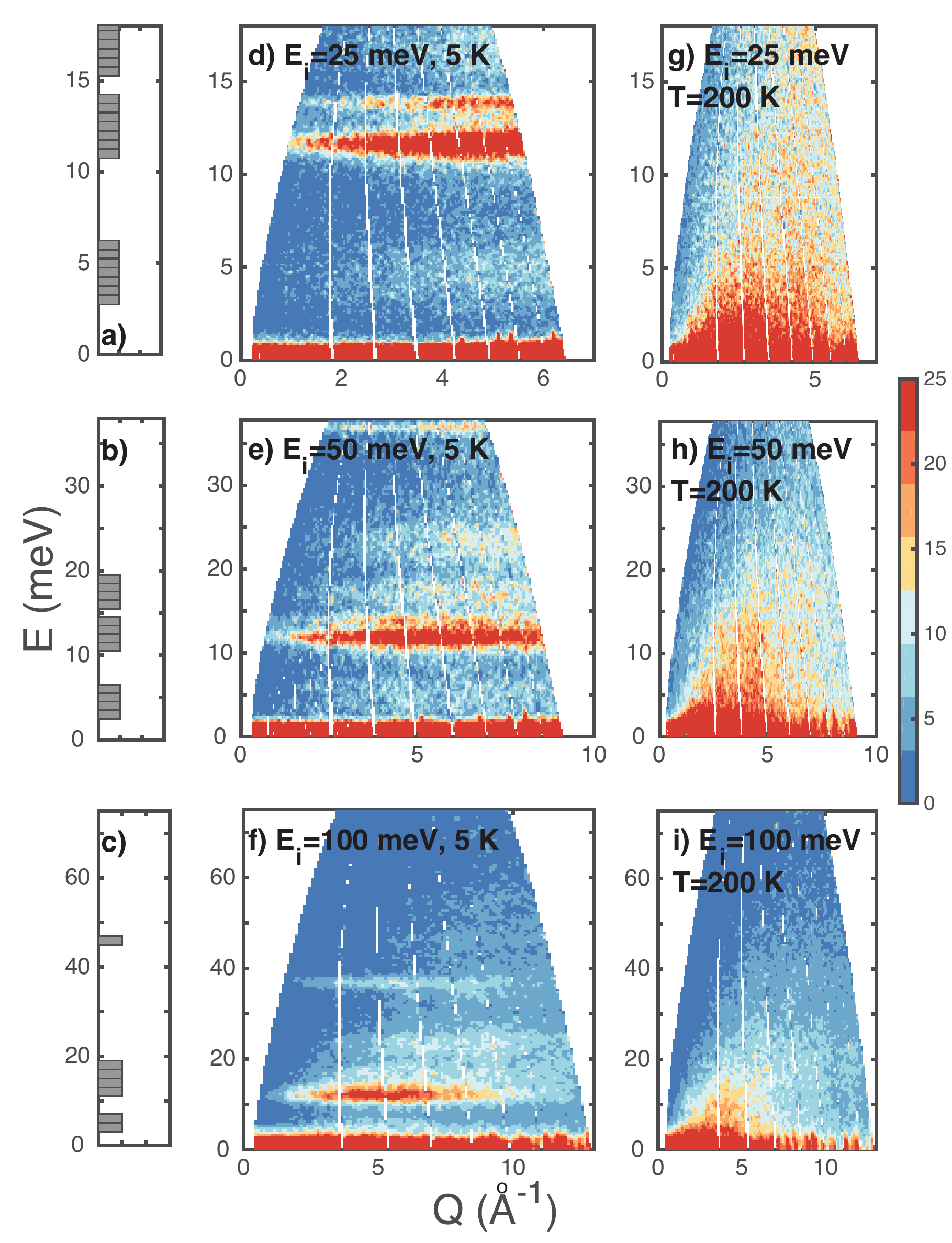}
\caption{\label{mari}  High-energy data taken on MARI in several configurations with different energy resolutions, compared to calculations.  $(a-c)$ CASTEP calculations showing where a finite phonon cross section exists.  $(d-f)$  MARI data taken at low temperatures in the orthorhombic phase.  $(g-i)$ shows MARI data taken in the tetragonal phase at 200~K.  The sharp, harmonic modes are replaced by quasistatic fluctuations extending up to high energies.}
\end{figure}

The MARI data at T=5 and 200 K is compared to Castep first-principles calculations in Fig. \ref{mari} $(a-c)$.  These calculations confirm the deuteration analysis (Fig. \ref{soft_mode}) that the low-energy mode that softens is associated with Pb-Br motions, and in particular octahedra tilting.  The result is also consistent with previous work on the phase transitions CsPbCl$_{3}$ which were found to result from soft tilt modes of the PbCl$_{6}$ octahedra.~\cite{Fujii74:9,Carabatos03:34,Shi98:29,Calistru97:82,supp}  The more intense, higher-energy modes that disappear at $T_{\rm c}$ are related to H motion.  CH$_{3}$NH$_{3}$PbBr$_{3}$ therefore represents a case where a displacive PbBr$_{6}$ octahedra transition is coupled to an order--disorder MA transition, corresponding to melting of the orientational order.  

We now discuss possible relations with electronic properties important for photoluminescence.  Measurements and calculations of the frequency-dependent dielectric response and electronic gap energies provide evidence that molecular vibrations are coupled to the electronic properties~\cite{Even14:118,Zheng15:6} and this has been further suggested by recent QENS results on the PbI$_{3}$ framework based compounds.~\cite{Chen:arxiv,Leguy15:6}   Further indications of the importance of the molecular vibrations come from the temperature dependent lifetime of the photoluminescence that shows an emission at 1.62 eV with a FWHM of 103 meV, sharpening at $\sim$ 150 K where molecular orientational order occurs.~\cite{Wehrenfennig14:5} While the energy scale of the molecular dynamics would be expected to be too low (less than meV displayed in Fig. \ref{figure1}) to influence the properties of the photoluminescence at this energy scale, the broad relaxational dynamics displayed in Fig. \ref{mari} do have a similar energy scale to the changes observed in photoluminescence.  While the molecular cation is not expected to directly contribute to the electronic band structure~\cite{Borriello08:77}, it is thought to indirectly influence the electronic band gap through coupling to the PbBr$_{3}$ framework.  As in many Pb perovksites, the lone electron pair on the Pb ion is strongly influential to the electronic band properties.~\cite{Payne06:96}  The strong coupling between the PbBr$_{3}$ framework and the MA molecular motions may also affect band structure details as predicted in Ref. \onlinecite{Zheng15:6}.

The potential impact on the electronic properties is further suggested in Ref. \onlinecite{Motta14:6} where it is noted that MA orientational driven indirect band gap processes may exist introducing a `dynamical band gap' discouraging carrier recombination.  The coupled relaxational dynamics observed here between the PbBr$_{3}$ framework and the MA molecule cation may provide a channel for such processes.  We note that a similar multiphonon process has been proposed for the luminescence in KCl.~\cite{Hlinka91:166}

In summary, we have reported a neutron inelastic scattering study of the coupled molecular and framework dynamics in CH$_{3}$NH$_{3}$PbBr$_{3}$.   We observe quasistatic molecular fluctuations coupled to a soft mode associated with tilting of the PbBr$_{3}$ framework.  The low temperature harmonic fluctuations cross over to fast overdamped relaxational dynamics at higher temperatures.  We have discussed possible coupling between these dynamics and the photoluminescence spectra.

\begin{acknowledgments}
 
This work was funded by the Carnegie Trust for the Universities of Scotland, the National Research Council of Canada, the Royal Society, EU-NMI3, STFC, and EPSRC.

\end{acknowledgments}


\begin{thebibliography}{42}%
\makeatletter
\providecommand \@ifxundefined [1]{%
 \@ifx{#1\undefined}
}%
\providecommand \@ifnum [1]{%
 \ifnum #1\expandafter \@firstoftwo
 \else \expandafter \@secondoftwo
 \fi
}%
\providecommand \@ifx [1]{%
 \ifx #1\expandafter \@firstoftwo
 \else \expandafter \@secondoftwo
 \fi
}%
\providecommand \natexlab [1]{#1}%
\providecommand \enquote  [1]{``#1''}%
\providecommand \bibnamefont  [1]{#1}%
\providecommand \bibfnamefont [1]{#1}%
\providecommand \citenamefont [1]{#1}%
\providecommand \href@noop [0]{\@secondoftwo}%
\providecommand \href [0]{\begingroup \@sanitize@url \@href}%
\providecommand \@href[1]{\@@startlink{#1}\@@href}%
\providecommand \@@href[1]{\endgroup#1\@@endlink}%
\providecommand \@sanitize@url [0]{\catcode `\\12\catcode `\$12\catcode
  `\&12\catcode `\#12\catcode `\^12\catcode `\_12\catcode `\%12\relax}%
\providecommand \@@startlink[1]{}%
\providecommand \@@endlink[0]{}%
\providecommand \url  [0]{\begingroup\@sanitize@url \@url }%
\providecommand \@url [1]{\endgroup\@href {#1}{\urlprefix }}%
\providecommand \urlprefix  [0]{URL }%
\providecommand \Eprint [0]{\href }%
\providecommand \doibase [0]{http://dx.doi.org/}%
\providecommand \selectlanguage [0]{\@gobble}%
\providecommand \bibinfo  [0]{\@secondoftwo}%
\providecommand \bibfield  [0]{\@secondoftwo}%
\providecommand \translation [1]{[#1]}%
\providecommand \BibitemOpen [0]{}%
\providecommand \bibitemStop [0]{}%
\providecommand \bibitemNoStop [0]{.\EOS\space}%
\providecommand \EOS [0]{\spacefactor3000\relax}%
\providecommand \BibitemShut  [1]{\csname bibitem#1\endcsname}%
\let\auto@bib@innerbib\@empty
\bibitem [{\citenamefont {Wyckoff}(1928)}]{Wyckoff28:1}%
  \BibitemOpen
  \bibfield  {author} {\bibinfo {author} {\bibfnamefont {R.}~\bibnamefont
  {Wyckoff}},\ }\href@noop {} {\bibfield  {journal} {\bibinfo  {journal} {Am.
  J. Sci.}\ ,\ \bibinfo {pages} {349}} (\bibinfo {year} {1928})}\BibitemShut
  {NoStop}%
\bibitem [{\citenamefont {Kojima}\ \emph {et~al.}(2009)\citenamefont {Kojima},
  \citenamefont {Teshima}, \citenamefont {Shirai},\ and\ \citenamefont
  {Miyasaka}}]{Kojima09:131}%
  \BibitemOpen
  \bibfield  {author} {\bibinfo {author} {\bibfnamefont {A.}~\bibnamefont
  {Kojima}}, \bibinfo {author} {\bibfnamefont {K.}~\bibnamefont {Teshima}},
  \bibinfo {author} {\bibfnamefont {Y.}~\bibnamefont {Shirai}}, \ and\ \bibinfo
  {author} {\bibfnamefont {T.}~\bibnamefont {Miyasaka}},\ }\href@noop {}
  {\bibfield  {journal} {\bibinfo  {journal} {J. Am. Chem. Soc.}\ }\textbf
  {\bibinfo {volume} {131}},\ \bibinfo {pages} {6050} (\bibinfo {year}
  {2009})}\BibitemShut {NoStop}%
\bibitem [{\citenamefont {Mitzi}\ \emph {et~al.}(2001)\citenamefont {Mitzi},
  \citenamefont {Chondroudis},\ and\ \citenamefont {Kagan}}]{Mitzi01:45}%
  \BibitemOpen
  \bibfield  {author} {\bibinfo {author} {\bibfnamefont {D.~B.}\ \bibnamefont
  {Mitzi}}, \bibinfo {author} {\bibfnamefont {K.}~\bibnamefont {Chondroudis}},
  \ and\ \bibinfo {author} {\bibfnamefont {C.~R.}\ \bibnamefont {Kagan}},\
  }\href@noop {} {\bibfield  {journal} {\bibinfo  {journal} {IBM J. Res.
  Devel.}\ }\textbf {\bibinfo {volume} {45}},\ \bibinfo {pages} {29} (\bibinfo
  {year} {2001})}\BibitemShut {NoStop}%
\bibitem [{\citenamefont {Mitzi}(2007)}]{Mitzi07:48}%
  \BibitemOpen
  \bibfield  {author} {\bibinfo {author} {\bibfnamefont {D.~B.}\ \bibnamefont
  {Mitzi}},\ }\href@noop {} {\bibfield  {journal} {\bibinfo  {journal} {Prog.
  Inorg. Chem.}\ }\textbf {\bibinfo {volume} {48}},\ \bibinfo {pages} {1}
  (\bibinfo {year} {2007})}\BibitemShut {NoStop}%
\bibitem [{\citenamefont {Park}(2013)}]{Park13:4}%
  \BibitemOpen
  \bibfield  {author} {\bibinfo {author} {\bibfnamefont {N.~G.}\ \bibnamefont
  {Park}},\ }\href@noop {} {\bibfield  {journal} {\bibinfo  {journal} {Phys.
  Chem. Letters}\ }\textbf {\bibinfo {volume} {4}},\ \bibinfo {pages} {2423}
  (\bibinfo {year} {2013})}\BibitemShut {NoStop}%
\bibitem [{\citenamefont {Swainson}\ \emph {et~al.}(2003)\citenamefont
  {Swainson}, \citenamefont {Hammond}, \citenamefont {Souliere}, \citenamefont
  {Knop},\ and\ \citenamefont {Massa}}]{Swainson03:176}%
  \BibitemOpen
  \bibfield  {author} {\bibinfo {author} {\bibfnamefont {I.~P.}\ \bibnamefont
  {Swainson}}, \bibinfo {author} {\bibfnamefont {R.~P.}\ \bibnamefont
  {Hammond}}, \bibinfo {author} {\bibfnamefont {C.}~\bibnamefont {Souliere}},
  \bibinfo {author} {\bibfnamefont {O.}~\bibnamefont {Knop}}, \ and\ \bibinfo
  {author} {\bibfnamefont {W.}~\bibnamefont {Massa}},\ }\href@noop {}
  {\bibfield  {journal} {\bibinfo  {journal} {J. Solid State Chem.}\ }\textbf
  {\bibinfo {volume} {176}},\ \bibinfo {pages} {97} (\bibinfo {year}
  {2003})}\BibitemShut {NoStop}%
\bibitem [{\citenamefont {Maalej}\ \emph {et~al.}(2015)\citenamefont {Maalej},
  \citenamefont {Bahri}, \citenamefont {Abid},\ and\ \citenamefont
  {Jaidane}}]{Ong03:176}%
  \BibitemOpen
  \bibfield  {author} {\bibinfo {author} {\bibfnamefont {A.}~\bibnamefont
  {Maalej}}, \bibinfo {author} {\bibfnamefont {M.}~\bibnamefont {Bahri}},
  \bibinfo {author} {\bibfnamefont {Y.}~\bibnamefont {Abid}}, \ and\ \bibinfo
  {author} {\bibfnamefont {N.}~\bibnamefont {Jaidane}},\ }\href@noop {}
  {\bibfield  {journal} {\bibinfo  {journal} {J. Phys. Chem. Lett.}\ }\textbf
  {\bibinfo {volume} {6}},\ \bibinfo {pages} {681} (\bibinfo {year}
  {2015})}\BibitemShut {NoStop}%
\bibitem [{\citenamefont {Futterer}\ \emph {et~al.}(1995)\citenamefont
  {Futterer}, \citenamefont {Withers}, \citenamefont {Welberry},\ and\
  \citenamefont {Depmeier}}]{Futterer95:7}%
  \BibitemOpen
  \bibfield  {author} {\bibinfo {author} {\bibfnamefont {K.}~\bibnamefont
  {Futterer}}, \bibinfo {author} {\bibfnamefont {R.~L.}\ \bibnamefont
  {Withers}}, \bibinfo {author} {\bibfnamefont {T.~R.}\ \bibnamefont
  {Welberry}}, \ and\ \bibinfo {author} {\bibfnamefont {W.}~\bibnamefont
  {Depmeier}},\ }\href@noop {} {\bibfield  {journal} {\bibinfo  {journal} {J.
  Phys.: Condens Matter}\ }\textbf {\bibinfo {volume} {7}},\ \bibinfo {pages}
  {4938} (\bibinfo {year} {1995})}\BibitemShut {NoStop}%
\bibitem [{\citenamefont {Knop}\ \emph {et~al.}(1990)\citenamefont {Knop},
  \citenamefont {Wasylishen}, \citenamefont {White}, \citenamefont {Cameron},\
  and\ \citenamefont {Oort}}]{Knop90:68}%
  \BibitemOpen
  \bibfield  {author} {\bibinfo {author} {\bibfnamefont {O.}~\bibnamefont
  {Knop}}, \bibinfo {author} {\bibfnamefont {R.~E.}\ \bibnamefont
  {Wasylishen}}, \bibinfo {author} {\bibfnamefont {M.~A.}\ \bibnamefont
  {White}}, \bibinfo {author} {\bibfnamefont {T.~S.}\ \bibnamefont {Cameron}},
  \ and\ \bibinfo {author} {\bibfnamefont {M.~J. M.~V.}\ \bibnamefont {Oort}},\
  }\href@noop {} {\bibfield  {journal} {\bibinfo  {journal} {Can. J. Chem.}\
  }\textbf {\bibinfo {volume} {68}},\ \bibinfo {pages} {412} (\bibinfo {year}
  {1990})}\BibitemShut {NoStop}%
\bibitem [{\citenamefont {Worhatch}\ \emph {et~al.}(2008)\citenamefont
  {Worhatch}, \citenamefont {Kim}, \citenamefont {Swainson}, \citenamefont
  {Yonkeu},\ and\ \citenamefont {Billinge}}]{Worhatch08:20}%
  \BibitemOpen
  \bibfield  {author} {\bibinfo {author} {\bibfnamefont {R.~J.}\ \bibnamefont
  {Worhatch}}, \bibinfo {author} {\bibfnamefont {H.}~\bibnamefont {Kim}},
  \bibinfo {author} {\bibfnamefont {I.}~\bibnamefont {Swainson}}, \bibinfo
  {author} {\bibfnamefont {A.~L.}\ \bibnamefont {Yonkeu}}, \ and\ \bibinfo
  {author} {\bibfnamefont {S.~J.}\ \bibnamefont {Billinge}},\ }\href@noop {}
  {\bibfield  {journal} {\bibinfo  {journal} {Chem. Mater}\ }\textbf {\bibinfo
  {volume} {20}},\ \bibinfo {pages} {1272} (\bibinfo {year}
  {2008})}\BibitemShut {NoStop}%
\bibitem [{\citenamefont {Kawamura}\ \emph {et~al.}(2002)\citenamefont
  {Kawamura}, \citenamefont {Mashiyama},\ and\ \citenamefont
  {Hasebe}}]{Mashiyama02:71}%
  \BibitemOpen
  \bibfield  {author} {\bibinfo {author} {\bibfnamefont {Y.}~\bibnamefont
  {Kawamura}}, \bibinfo {author} {\bibfnamefont {H.}~\bibnamefont {Mashiyama}},
  \ and\ \bibinfo {author} {\bibfnamefont {K.}~\bibnamefont {Hasebe}},\
  }\href@noop {} {\bibfield  {journal} {\bibinfo  {journal} {J Korean Phys.
  Soc.}\ }\textbf {\bibinfo {volume} {71}},\ \bibinfo {pages} {1694} (\bibinfo
  {year} {2002})}\BibitemShut {NoStop}%
\bibitem [{\citenamefont {Mashiyama}\ \emph
  {et~al.}(2007{\natexlab{a}})\citenamefont {Mashiyama}, \citenamefont
  {Kawamura}, \citenamefont {Kasano}, \citenamefont {Asahi}, \citenamefont
  {Noda},\ and\ \citenamefont {Kimura}}]{Mashiyama07:348}%
  \BibitemOpen
  \bibfield  {author} {\bibinfo {author} {\bibfnamefont {H.}~\bibnamefont
  {Mashiyama}}, \bibinfo {author} {\bibfnamefont {Y.}~\bibnamefont {Kawamura}},
  \bibinfo {author} {\bibfnamefont {H.}~\bibnamefont {Kasano}}, \bibinfo
  {author} {\bibfnamefont {T.}~\bibnamefont {Asahi}}, \bibinfo {author}
  {\bibfnamefont {Y.}~\bibnamefont {Noda}}, \ and\ \bibinfo {author}
  {\bibfnamefont {H.}~\bibnamefont {Kimura}},\ }\href@noop {} {\bibfield
  {journal} {\bibinfo  {journal} {Ferroelectrics}\ }\textbf {\bibinfo {volume}
  {348}},\ \bibinfo {pages} {182} (\bibinfo {year}
  {2007}{\natexlab{a}})}\BibitemShut {NoStop}%
\bibitem [{\citenamefont {Chi}\ \emph {et~al.}(2005)\citenamefont {Chi},
  \citenamefont {Swainson}, \citenamefont {Cranswick}, \citenamefont {Her},
  \citenamefont {Stephens},\ and\ \citenamefont {Knop}}]{Chi05:178}%
  \BibitemOpen
  \bibfield  {author} {\bibinfo {author} {\bibfnamefont {L.}~\bibnamefont
  {Chi}}, \bibinfo {author} {\bibfnamefont {I.~P.}\ \bibnamefont {Swainson}},
  \bibinfo {author} {\bibfnamefont {L.}~\bibnamefont {Cranswick}}, \bibinfo
  {author} {\bibfnamefont {J.~H.}\ \bibnamefont {Her}}, \bibinfo {author}
  {\bibfnamefont {P.}~\bibnamefont {Stephens}}, \ and\ \bibinfo {author}
  {\bibfnamefont {O.}~\bibnamefont {Knop}},\ }\href@noop {} {\bibfield
  {journal} {\bibinfo  {journal} {J. Solid State Chem.}\ }\textbf {\bibinfo
  {volume} {178}},\ \bibinfo {pages} {1376} (\bibinfo {year}
  {2005})}\BibitemShut {NoStop}%
\bibitem [{\citenamefont {Mashiyama}\ \emph
  {et~al.}(2007{\natexlab{b}})\citenamefont {Mashiyama}, \citenamefont
  {Kawamura},\ and\ \citenamefont {Kubota}}]{Mashiyama07:51}%
  \BibitemOpen
  \bibfield  {author} {\bibinfo {author} {\bibfnamefont {H.}~\bibnamefont
  {Mashiyama}}, \bibinfo {author} {\bibfnamefont {Y.}~\bibnamefont {Kawamura}},
  \ and\ \bibinfo {author} {\bibfnamefont {Y.}~\bibnamefont {Kubota}},\
  }\href@noop {} {\bibfield  {journal} {\bibinfo  {journal} {J Korean Phys.
  Soc.}\ }\textbf {\bibinfo {volume} {51}},\ \bibinfo {pages} {850} (\bibinfo
  {year} {2007}{\natexlab{b}})}\BibitemShut {NoStop}%
\bibitem [{\citenamefont {Quarti}\ \emph {et~al.}(2014)\citenamefont {Quarti},
  \citenamefont {Mosconi},\ and\ \citenamefont {Angelis}}]{Quarti14:26}%
  \BibitemOpen
  \bibfield  {author} {\bibinfo {author} {\bibfnamefont {C.}~\bibnamefont
  {Quarti}}, \bibinfo {author} {\bibfnamefont {E.}~\bibnamefont {Mosconi}}, \
  and\ \bibinfo {author} {\bibfnamefont {F.~D.}\ \bibnamefont {Angelis}},\
  }\href@noop {} {\bibfield  {journal} {\bibinfo  {journal} {Chem. Mater.}\
  }\textbf {\bibinfo {volume} {26}},\ \bibinfo {pages} {6557} (\bibinfo {year}
  {2014})}\BibitemShut {NoStop}%
\bibitem [{\citenamefont {Zheng}\ \emph {et~al.}(2015)\citenamefont {Zheng},
  \citenamefont {Takenaka}, \citenamefont {Wang}, \citenamefont {Koocher},\
  and\ \citenamefont {Rappe}}]{Zheng15:6}%
  \BibitemOpen
  \bibfield  {author} {\bibinfo {author} {\bibfnamefont {F.}~\bibnamefont
  {Zheng}}, \bibinfo {author} {\bibfnamefont {H.}~\bibnamefont {Takenaka}},
  \bibinfo {author} {\bibfnamefont {F.}~\bibnamefont {Wang}}, \bibinfo {author}
  {\bibfnamefont {N.~Z.}\ \bibnamefont {Koocher}}, \ and\ \bibinfo {author}
  {\bibfnamefont {A.~M.}\ \bibnamefont {Rappe}},\ }\href@noop {} {\bibfield
  {journal} {\bibinfo  {journal} {J. Phys. Chem. Lett.}\ }\textbf {\bibinfo
  {volume} {6}},\ \bibinfo {pages} {31} (\bibinfo {year} {2015})}\BibitemShut
  {NoStop}%
\bibitem [{\citenamefont {Segall}\ \emph {et~al.}(2002)\citenamefont {Segall},
  \citenamefont {Lindan}, \citenamefont {Probert}, \citenamefont {Pickard},
  \citenamefont {Hasnip}, \citenamefont {Clark},\ and\ \citenamefont
  {Payne}}]{CASTEP}%
  \BibitemOpen
  \bibfield  {author} {\bibinfo {author} {\bibfnamefont {M.}~\bibnamefont
  {Segall}}, \bibinfo {author} {\bibfnamefont {P.}~\bibnamefont {Lindan}},
  \bibinfo {author} {\bibfnamefont {M.}~\bibnamefont {Probert}}, \bibinfo
  {author} {\bibfnamefont {C.}~\bibnamefont {Pickard}}, \bibinfo {author}
  {\bibfnamefont {P.}~\bibnamefont {Hasnip}}, \bibinfo {author} {\bibfnamefont
  {S.}~\bibnamefont {Clark}}, \ and\ \bibinfo {author} {\bibfnamefont
  {M.}~\bibnamefont {Payne}},\ }\href@noop {} {\bibfield  {journal} {\bibinfo
  {journal} {J. Phys. Cond. Matter}\ }\textbf {\bibinfo {volume} {14}},\
  \bibinfo {pages} {2717} (\bibinfo {year} {2002})}\BibitemShut {NoStop}%
\bibitem [{\citenamefont {Ramirez-Cuesta}(2004)}]{ACLIMAX}%
  \BibitemOpen
  \bibfield  {author} {\bibinfo {author} {\bibfnamefont {A.}~\bibnamefont
  {Ramirez-Cuesta}},\ }\href@noop {} {\bibfield  {journal} {\bibinfo  {journal}
  {Comput. Phys. Commun.}\ }\textbf {\bibinfo {volume} {157}},\ \bibinfo
  {pages} {226} (\bibinfo {year} {2004})}\BibitemShut {NoStop}%
\bibitem [{sup()}]{supp}%
  \BibitemOpen
  \href@noop {} {\bibinfo  {journal} {See Supplemental Material at [*WEB
  ADDRESS*] for additional information regarding analysis, mode assignment, and
  open access data.}\ }\BibitemShut {NoStop}%
\bibitem [{\citenamefont {Kawamura}\ and\ \citenamefont
  {Mashiyama}(1999)}]{Kawamura99:35}%
  \BibitemOpen
\bibfield  {journal} {  }\bibfield  {author} {\bibinfo {author} {\bibfnamefont
  {Y.}~\bibnamefont {Kawamura}}\ and\ \bibinfo {author} {\bibfnamefont
  {H.}~\bibnamefont {Mashiyama}},\ }\href@noop {} {\bibfield  {journal}
  {\bibinfo  {journal} {J Korean Phys. Soc.}\ }\textbf {\bibinfo {volume}
  {35}},\ \bibinfo {pages} {S1437} (\bibinfo {year} {1999})}\BibitemShut
  {NoStop}%
\bibitem [{\citenamefont {Swainson}(2005)}]{Swainson05:61}%
  \BibitemOpen
  \bibfield  {author} {\bibinfo {author} {\bibfnamefont {I.~P.}\ \bibnamefont
  {Swainson}},\ }\href@noop {} {\bibfield  {journal} {\bibinfo  {journal} {Acta
  Crystallogr. B}\ }\textbf {\bibinfo {volume} {61}},\ \bibinfo {pages} {616}
  (\bibinfo {year} {2005})}\BibitemShut {NoStop}%
\bibitem [{\citenamefont {Line}\ \emph {et~al.}(1994)\citenamefont {Line},
  \citenamefont {Winkler},\ and\ \citenamefont {Dove}}]{Line94:21}%
  \BibitemOpen
  \bibfield  {author} {\bibinfo {author} {\bibfnamefont {C.~M.~B.}\
  \bibnamefont {Line}}, \bibinfo {author} {\bibfnamefont {B.}~\bibnamefont
  {Winkler}}, \ and\ \bibinfo {author} {\bibfnamefont {M.~T.}\ \bibnamefont
  {Dove}},\ }\href@noop {} {\bibfield  {journal} {\bibinfo  {journal} {Phys.
  Chem. Minerals}\ }\textbf {\bibinfo {volume} {21}},\ \bibinfo {pages} {451}
  (\bibinfo {year} {1994})}\BibitemShut {NoStop}%
\bibitem [{\citenamefont {Bee}(1988)}]{Bee:book}%
  \BibitemOpen
  \bibfield  {author} {\bibinfo {author} {\bibfnamefont {M.}~\bibnamefont
  {Bee}},\ }\href@noop {} {\emph {\bibinfo {title} {Quasielastic neutron
  scattering:principles and applications in solid state chemistry, biology, and
  materials science}}}\ (\bibinfo  {publisher} {Adam Hilger},\ \bibinfo
  {address} {Bristol},\ \bibinfo {year} {1988})\BibitemShut {NoStop}%
\bibitem [{\citenamefont {Kieslich}\ \emph {et~al.}(2014)\citenamefont
  {Kieslich}, \citenamefont {Sun},\ and\ \citenamefont
  {Cheetham}}]{Kieslich14:5}%
  \BibitemOpen
  \bibfield  {author} {\bibinfo {author} {\bibfnamefont {G.}~\bibnamefont
  {Kieslich}}, \bibinfo {author} {\bibfnamefont {S.}~\bibnamefont {Sun}}, \
  and\ \bibinfo {author} {\bibfnamefont {A.~K.}\ \bibnamefont {Cheetham}},\
  }\href@noop {} {\bibfield  {journal} {\bibinfo  {journal} {Chem. Sci.}\
  }\textbf {\bibinfo {volume} {5}},\ \bibinfo {pages} {4712} (\bibinfo {year}
  {2014})}\BibitemShut {NoStop}%
\bibitem [{\citenamefont {Varma}\ \emph {et~al.}(1992)\citenamefont {Varma},
  \citenamefont {Bhattacharjee}, \citenamefont {Vasan},\ and\ \citenamefont
  {Rao}}]{Varma92:48}%
  \BibitemOpen
  \bibfield  {author} {\bibinfo {author} {\bibfnamefont {Y.}~\bibnamefont
  {Varma}}, \bibinfo {author} {\bibfnamefont {R.}~\bibnamefont
  {Bhattacharjee}}, \bibinfo {author} {\bibfnamefont {H.~N.}\ \bibnamefont
  {Vasan}}, \ and\ \bibinfo {author} {\bibfnamefont {C.~N.~R.}\ \bibnamefont
  {Rao}},\ }\href@noop {} {\bibfield  {journal} {\bibinfo  {journal}
  {Spectrochimica Acta}\ }\textbf {\bibinfo {volume} {48A}},\ \bibinfo {pages}
  {1631} (\bibinfo {year} {1992})}\BibitemShut {NoStop}%
\bibitem [{\citenamefont {Trevino}\ and\ \citenamefont
  {Rymes}(1980)}]{Trevino80:73}%
  \BibitemOpen
  \bibfield  {author} {\bibinfo {author} {\bibfnamefont {S.~F.}\ \bibnamefont
  {Trevino}}\ and\ \bibinfo {author} {\bibfnamefont {W.~H.}\ \bibnamefont
  {Rymes}},\ }\href@noop {} {\bibfield  {journal} {\bibinfo  {journal} {J.
  Chem. Phys.}\ }\textbf {\bibinfo {volume} {73}},\ \bibinfo {pages} {3001}
  (\bibinfo {year} {1980})}\BibitemShut {NoStop}%
\bibitem [{\citenamefont {Choi}\ \emph {et~al.}(2013)\citenamefont {Choi},
  \citenamefont {Yang}, \citenamefont {Norman}, \citenamefont {Billinge},\ and\
  \citenamefont {Owen}}]{Choi13:14}%
  \BibitemOpen
  \bibfield  {author} {\bibinfo {author} {\bibfnamefont {J.}~\bibnamefont
  {Choi}}, \bibinfo {author} {\bibfnamefont {X.}~\bibnamefont {Yang}}, \bibinfo
  {author} {\bibfnamefont {Z.}~\bibnamefont {Norman}}, \bibinfo {author}
  {\bibfnamefont {S.}~\bibnamefont {Billinge}}, \ and\ \bibinfo {author}
  {\bibfnamefont {J.}~\bibnamefont {Owen}},\ }\href@noop {} {\bibfield
  {journal} {\bibinfo  {journal} {Nano. Lett.}\ }\textbf {\bibinfo {volume}
  {14}},\ \bibinfo {pages} {127} (\bibinfo {year} {2013})}\BibitemShut
  {NoStop}%
\bibitem [{\citenamefont {Cochran}(1960)}]{Cochran60:9}%
  \BibitemOpen
  \bibfield  {author} {\bibinfo {author} {\bibfnamefont {W.}~\bibnamefont
  {Cochran}},\ }\href@noop {} {\bibfield  {journal} {\bibinfo  {journal} {Adv.
  in Physics}\ }\textbf {\bibinfo {volume} {9}},\ \bibinfo {pages} {387}
  (\bibinfo {year} {1960})}\BibitemShut {NoStop}%
\bibitem [{\citenamefont {Gesi}\ \emph {et~al.}(1972)\citenamefont {Gesi},
  \citenamefont {Axe}, \citenamefont {Shirane},\ and\ \citenamefont
  {Linz}}]{Gesi72:5}%
  \BibitemOpen
  \bibfield  {author} {\bibinfo {author} {\bibfnamefont {K.}~\bibnamefont
  {Gesi}}, \bibinfo {author} {\bibfnamefont {J.~D.}\ \bibnamefont {Axe}},
  \bibinfo {author} {\bibfnamefont {G.}~\bibnamefont {Shirane}}, \ and\
  \bibinfo {author} {\bibfnamefont {A.}~\bibnamefont {Linz}},\ }\href@noop {}
  {\bibfield  {journal} {\bibinfo  {journal} {Phys. Rev. B}\ }\textbf {\bibinfo
  {volume} {5}},\ \bibinfo {pages} {1933} (\bibinfo {year} {1972})}\BibitemShut
  {NoStop}%
\bibitem [{\citenamefont {Shirane}\ and\ \citenamefont
  {Yamada}(1969)}]{Shirane177:69}%
  \BibitemOpen
  \bibfield  {author} {\bibinfo {author} {\bibfnamefont {G.}~\bibnamefont
  {Shirane}}\ and\ \bibinfo {author} {\bibfnamefont {Y.}~\bibnamefont
  {Yamada}},\ }\href@noop {} {\bibfield  {journal} {\bibinfo  {journal} {Phys.
  Rev.}\ }\textbf {\bibinfo {volume} {177}},\ \bibinfo {pages} {858} (\bibinfo
  {year} {1969})}\BibitemShut {NoStop}%
\bibitem [{\citenamefont {Fujii}\ \emph {et~al.}(1974)\citenamefont {Fujii},
  \citenamefont {Hoshino}, \citenamefont {Yamada},\ and\ \citenamefont
  {Shirane}}]{Fujii74:9}%
  \BibitemOpen
  \bibfield  {author} {\bibinfo {author} {\bibfnamefont {Y.}~\bibnamefont
  {Fujii}}, \bibinfo {author} {\bibfnamefont {S.}~\bibnamefont {Hoshino}},
  \bibinfo {author} {\bibfnamefont {Y.}~\bibnamefont {Yamada}}, \ and\ \bibinfo
  {author} {\bibfnamefont {G.}~\bibnamefont {Shirane}},\ }\href@noop {}
  {\bibfield  {journal} {\bibinfo  {journal} {Phys. Rev. B}\ }\textbf {\bibinfo
  {volume} {9}},\ \bibinfo {pages} {4549} (\bibinfo {year} {1974})}\BibitemShut
  {NoStop}%
\bibitem [{\citenamefont {Carabatos-Nedelec}\ \emph {et~al.}(2003)\citenamefont
  {Carabatos-Nedelec}, \citenamefont {Oussaid},\ and\ \citenamefont
  {Nitsch}}]{Carabatos03:34}%
  \BibitemOpen
  \bibfield  {author} {\bibinfo {author} {\bibfnamefont {C.}~\bibnamefont
  {Carabatos-Nedelec}}, \bibinfo {author} {\bibfnamefont {M.}~\bibnamefont
  {Oussaid}}, \ and\ \bibinfo {author} {\bibfnamefont {K.}~\bibnamefont
  {Nitsch}},\ }\href@noop {} {\bibfield  {journal} {\bibinfo  {journal} {J.
  Raman Spectrosc.}\ }\textbf {\bibinfo {volume} {34}},\ \bibinfo {pages} {388}
  (\bibinfo {year} {2003})}\BibitemShut {NoStop}%
\bibitem [{\citenamefont {Shi}\ \emph {et~al.}(1998)\citenamefont {Shi},
  \citenamefont {Kume}, \citenamefont {Pelzl}, \citenamefont {Xu},\ and\
  \citenamefont {Wu}}]{Shi98:29}%
  \BibitemOpen
  \bibfield  {author} {\bibinfo {author} {\bibfnamefont {J.~R.}\ \bibnamefont
  {Shi}}, \bibinfo {author} {\bibfnamefont {Y.}~\bibnamefont {Kume}}, \bibinfo
  {author} {\bibfnamefont {J.}~\bibnamefont {Pelzl}}, \bibinfo {author}
  {\bibfnamefont {Y.~C.}\ \bibnamefont {Xu}}, \ and\ \bibinfo {author}
  {\bibfnamefont {X.}~\bibnamefont {Wu}},\ }\href@noop {} {\bibfield  {journal}
  {\bibinfo  {journal} {J. Raman Spectrosc.}\ }\textbf {\bibinfo {volume}
  {29}},\ \bibinfo {pages} {149} (\bibinfo {year} {1998})}\BibitemShut
  {NoStop}%
\bibitem [{\citenamefont {Calistru}\ \emph {et~al.}(1997)\citenamefont
  {Calistru}, \citenamefont {Mihut}, \citenamefont {Lefrant},\ and\
  \citenamefont {Baltog}}]{Calistru97:82}%
  \BibitemOpen
  \bibfield  {author} {\bibinfo {author} {\bibfnamefont {D.~M.}\ \bibnamefont
  {Calistru}}, \bibinfo {author} {\bibfnamefont {L.}~\bibnamefont {Mihut}},
  \bibinfo {author} {\bibfnamefont {S.}~\bibnamefont {Lefrant}}, \ and\
  \bibinfo {author} {\bibfnamefont {I.}~\bibnamefont {Baltog}},\ }\href@noop {}
  {\bibfield  {journal} {\bibinfo  {journal} {J. Appl. Phys.}\ }\textbf
  {\bibinfo {volume} {82}},\ \bibinfo {pages} {5391} (\bibinfo {year}
  {1997})}\BibitemShut {NoStop}%
\bibitem [{\citenamefont {Even}\ \emph {et~al.}(2014)\citenamefont {Even},
  \citenamefont {Pedesseau},\ and\ \citenamefont {Katan}}]{Even14:118}%
  \BibitemOpen
  \bibfield  {author} {\bibinfo {author} {\bibfnamefont {J.}~\bibnamefont
  {Even}}, \bibinfo {author} {\bibfnamefont {L.}~\bibnamefont {Pedesseau}}, \
  and\ \bibinfo {author} {\bibfnamefont {C.}~\bibnamefont {Katan}},\
  }\href@noop {} {\bibfield  {journal} {\bibinfo  {journal} {J. Phys. Chem. C}\
  }\textbf {\bibinfo {volume} {118}},\ \bibinfo {pages} {11566} (\bibinfo
  {year} {2014})}\BibitemShut {NoStop}%
\bibitem [{\citenamefont {Chen}\ \emph {et~al.}()\citenamefont {Chen},
  \citenamefont {Foley}, \citenamefont {Ipek}, \citenamefont {Tyagi},
  \citenamefont {Copley}, \citenamefont {Brown}, \citenamefont {Choi},\ and\
  \citenamefont {Lee}}]{Chen:arxiv}%
  \BibitemOpen
  \bibfield  {author} {\bibinfo {author} {\bibfnamefont {T.}~\bibnamefont
  {Chen}}, \bibinfo {author} {\bibfnamefont {J.}~\bibnamefont {Foley}},
  \bibinfo {author} {\bibfnamefont {B.}~\bibnamefont {Ipek}}, \bibinfo {author}
  {\bibfnamefont {M.}~\bibnamefont {Tyagi}}, \bibinfo {author} {\bibfnamefont
  {J.~R.~D.}\ \bibnamefont {Copley}}, \bibinfo {author} {\bibfnamefont {C.~M.}\
  \bibnamefont {Brown}}, \bibinfo {author} {\bibfnamefont {J.~J.}\ \bibnamefont
  {Choi}}, \ and\ \bibinfo {author} {\bibfnamefont {S.~H.}\ \bibnamefont
  {Lee}},\ }\href@noop {} {\bibinfo  {journal} {unpublished
  (arXiv:1506.02205)}\ }\BibitemShut {NoStop}%
\bibitem [{\citenamefont {Leguy}\ \emph {et~al.}(2015)\citenamefont {Leguy},
  \citenamefont {Frost}, \citenamefont {McMahon}, \citenamefont {Sakai},
  \citenamefont {Kockelmann}, \citenamefont {Law}, \citenamefont {Li},
  \citenamefont {Foglia}, \citenamefont {Walsh}, \citenamefont {O'Reagan},
  \citenamefont {Nelson}, \citenamefont {Cabral},\ and\ \citenamefont
  {Barnes}}]{Leguy15:6}%
  \BibitemOpen
\bibfield  {journal} {  }\bibfield  {author} {\bibinfo {author} {\bibfnamefont
  {A.~M.~A.}\ \bibnamefont {Leguy}}, \bibinfo {author} {\bibfnamefont {J.~M.}\
  \bibnamefont {Frost}}, \bibinfo {author} {\bibfnamefont {A.~P.}\ \bibnamefont
  {McMahon}}, \bibinfo {author} {\bibfnamefont {V.~G.}\ \bibnamefont {Sakai}},
  \bibinfo {author} {\bibfnamefont {W.}~\bibnamefont {Kockelmann}}, \bibinfo
  {author} {\bibfnamefont {C.}~\bibnamefont {Law}}, \bibinfo {author}
  {\bibfnamefont {X.}~\bibnamefont {Li}}, \bibinfo {author} {\bibfnamefont
  {F.}~\bibnamefont {Foglia}}, \bibinfo {author} {\bibfnamefont
  {A.}~\bibnamefont {Walsh}}, \bibinfo {author} {\bibfnamefont {B.~C.}\
  \bibnamefont {O'Reagan}}, \bibinfo {author} {\bibfnamefont {J.}~\bibnamefont
  {Nelson}}, \bibinfo {author} {\bibfnamefont {J.~T.}\ \bibnamefont {Cabral}},
  \ and\ \bibinfo {author} {\bibfnamefont {P.~R.~F.}\ \bibnamefont {Barnes}},\
  }\href@noop {} {\bibfield  {journal} {\bibinfo  {journal} {Nat. Comm.}\
  }\textbf {\bibinfo {volume} {6}},\ \bibinfo {pages} {7124} (\bibinfo {year}
  {2015})}\BibitemShut {NoStop}%
\bibitem [{\citenamefont {Wehrenfennig}\ \emph {et~al.}(2014)\citenamefont
  {Wehrenfennig}, \citenamefont {Liu}, \citenamefont {Snaith}, \citenamefont
  {Johnston},\ and\ \citenamefont {Herz}}]{Wehrenfennig14:5}%
  \BibitemOpen
  \bibfield  {author} {\bibinfo {author} {\bibfnamefont {C.}~\bibnamefont
  {Wehrenfennig}}, \bibinfo {author} {\bibfnamefont {M.}~\bibnamefont {Liu}},
  \bibinfo {author} {\bibfnamefont {H.~J.}\ \bibnamefont {Snaith}}, \bibinfo
  {author} {\bibfnamefont {M.~B.}\ \bibnamefont {Johnston}}, \ and\ \bibinfo
  {author} {\bibfnamefont {L.~M.}\ \bibnamefont {Herz}},\ }\href@noop {}
  {\bibfield  {journal} {\bibinfo  {journal} {J. Phys. Chem. Lett.}\ }\textbf
  {\bibinfo {volume} {5}},\ \bibinfo {pages} {1300} (\bibinfo {year}
  {2014})}\BibitemShut {NoStop}%
\bibitem [{\citenamefont {Borriello}\ \emph {et~al.}(2008)\citenamefont
  {Borriello}, \citenamefont {Cantele},\ and\ \citenamefont
  {Ninno}}]{Borriello08:77}%
  \BibitemOpen
  \bibfield  {author} {\bibinfo {author} {\bibfnamefont {I.}~\bibnamefont
  {Borriello}}, \bibinfo {author} {\bibfnamefont {G.}~\bibnamefont {Cantele}},
  \ and\ \bibinfo {author} {\bibfnamefont {D.}~\bibnamefont {Ninno}},\
  }\href@noop {} {\bibfield  {journal} {\bibinfo  {journal} {Phys. Rev. B}\
  }\textbf {\bibinfo {volume} {77}},\ \bibinfo {pages} {235214} (\bibinfo
  {year} {2008})}\BibitemShut {NoStop}%
\bibitem [{\citenamefont {Payne}\ \emph {et~al.}(2006)\citenamefont {Payne},
  \citenamefont {Egdell}, \citenamefont {Walsh}, \citenamefont {Watson},
  \citenamefont {Guo}, \citenamefont {Glans}, \citenamefont {Learmonth},\ and\
  \citenamefont {Smith}}]{Payne06:96}%
  \BibitemOpen
  \bibfield  {author} {\bibinfo {author} {\bibfnamefont {D.~J.}\ \bibnamefont
  {Payne}}, \bibinfo {author} {\bibfnamefont {R.~G.}\ \bibnamefont {Egdell}},
  \bibinfo {author} {\bibfnamefont {A.}~\bibnamefont {Walsh}}, \bibinfo
  {author} {\bibfnamefont {G.~W.}\ \bibnamefont {Watson}}, \bibinfo {author}
  {\bibfnamefont {J.}~\bibnamefont {Guo}}, \bibinfo {author} {\bibfnamefont
  {P.~A.}\ \bibnamefont {Glans}}, \bibinfo {author} {\bibfnamefont
  {T.}~\bibnamefont {Learmonth}}, \ and\ \bibinfo {author} {\bibfnamefont
  {K.~E.}\ \bibnamefont {Smith}},\ }\href@noop {} {\bibfield  {journal}
  {\bibinfo  {journal} {Phys. Rev. Lett.}\ }\textbf {\bibinfo {volume} {96}},\
  \bibinfo {pages} {157403} (\bibinfo {year} {2006})}\BibitemShut {NoStop}%
\bibitem [{\citenamefont {Motta}\ \emph {et~al.}(2015)\citenamefont {Motta},
  \citenamefont {El-Mellouhi}, \citenamefont {Kais}, \citenamefont {Tabet},
  \citenamefont {Alharbi},\ and\ \citenamefont {Sanvito}}]{Motta14:6}%
  \BibitemOpen
  \bibfield  {author} {\bibinfo {author} {\bibfnamefont {C.}~\bibnamefont
  {Motta}}, \bibinfo {author} {\bibfnamefont {F.}~\bibnamefont {El-Mellouhi}},
  \bibinfo {author} {\bibfnamefont {S.}~\bibnamefont {Kais}}, \bibinfo {author}
  {\bibfnamefont {N.}~\bibnamefont {Tabet}}, \bibinfo {author} {\bibfnamefont
  {F.}~\bibnamefont {Alharbi}}, \ and\ \bibinfo {author} {\bibfnamefont
  {S.}~\bibnamefont {Sanvito}},\ }\href@noop {} {\bibfield  {journal} {\bibinfo
   {journal} {Nat. Comm.}\ }\textbf {\bibinfo {volume} {6}},\ \bibinfo {pages}
  {7026} (\bibinfo {year} {2015})}\BibitemShut {NoStop}%
\bibitem [{\citenamefont {Hlinka}\ \emph {et~al.}(1991)\citenamefont {Hlinka},
  \citenamefont {Mihokova},\ and\ \citenamefont {Nikl}}]{Hlinka91:166}%
  \BibitemOpen
  \bibfield  {author} {\bibinfo {author} {\bibfnamefont {J.}~\bibnamefont
  {Hlinka}}, \bibinfo {author} {\bibfnamefont {E.}~\bibnamefont {Mihokova}}, \
  and\ \bibinfo {author} {\bibfnamefont {M.}~\bibnamefont {Nikl}},\ }\href@noop
  {} {\bibfield  {journal} {\bibinfo  {journal} {Phys. Stat. Sol. (b)}\
  }\textbf {\bibinfo {volume} {166}},\ \bibinfo {pages} {503} (\bibinfo {year}
  {1991})}\BibitemShut {NoStop}%
\end{thebibliography}

%

\end{document}


\title{Supplementary information for ``From soft harmonic phonons to fast relaxational dynamics in CH$_{3}$NH$_{3}$PbBr$_{3}$"}

\author{I.P. Swainson}
\affiliation{National Research Council, Chalk River, Ontario, KOJ 1JO, Canada}

\author{C. Stock}
\affiliation{School of Physics and Astronomy, University of Edinburgh, Edinburgh EH9 3JZ, UK}

\author{S. F. Parker}
\affiliation{ISIS Facility, Rutherford Appleton Laboratory, Chilton, Didcot OX11 0QX, UK}

\author{L. Van Eijck}
\affiliation{Institut Laue-Langevin, 6 rue Jules Horowitz, Boite Postale 156, 38042 Grenoble Cedex 9, France}
\affiliation{Reactor Institute Delft, Delft University of Technology, 2629JB Delft, Netherlands}

\author{M. Russina}
\affiliation{Helmholtz Zentrum Berlin fur Materialien und Energie GmbH, D-14109 Berlin, Germany}

\author{J. W. Taylor}
\affiliation{ISIS Facility, Rutherford Appleton Laboratory, Chilton, Didcot OX11 0QX, UK}
\affiliation{European Spallation Source, Universitetparken 1, 2100, Copenhagen, Denmark}

\date{\today}

\begin{abstract} 

Supplementary information is provided in support of the main text.  Detailed information on the experimental setup (I), analysis of the ``fixed window" elastic scan done on IN10 (II), and a discussion of the higher energy modes (III) probed on the MARI direct geometry spectrometer.  In particular, we compare the higher energy lattice vibration modes taken at low temperatures with previously studied OIP's using Raman spectroscopy.

\end{abstract}

\pacs{}

\maketitle

\section{Experimental details}

Given the broad dynamic range required to probe both the molecular excitations and the lattice modes of the PbBr$_{6}$ host lattice, several instruments with overlapping energy resolutions were required.  We first performed a series of experiments on the IN10 backscattering instrument (ILL, France).  IN10 consists of a large, vibrating Si(111) monochromator that Doppler shifts incident neutrons with energies near E$_{i}$=2.08 meV and directs them onto the sample.  Neutrons scattered from the sample are then energy analyzed by a large bank of Si(111) crystals, which backscatter neutrons having energy E$_{f}$=2.08 meV through the sample and onto a series detectors.  The dynamic range of this spectrometer is $\pm$ 10 $\mu eV$, and the elastic resolution is $\delta E$=0.5 $\mu eV$ (half width).

\renewcommand{\thefigure}{S1}
\begin{figure}[t]
\includegraphics[width=8.5cm] {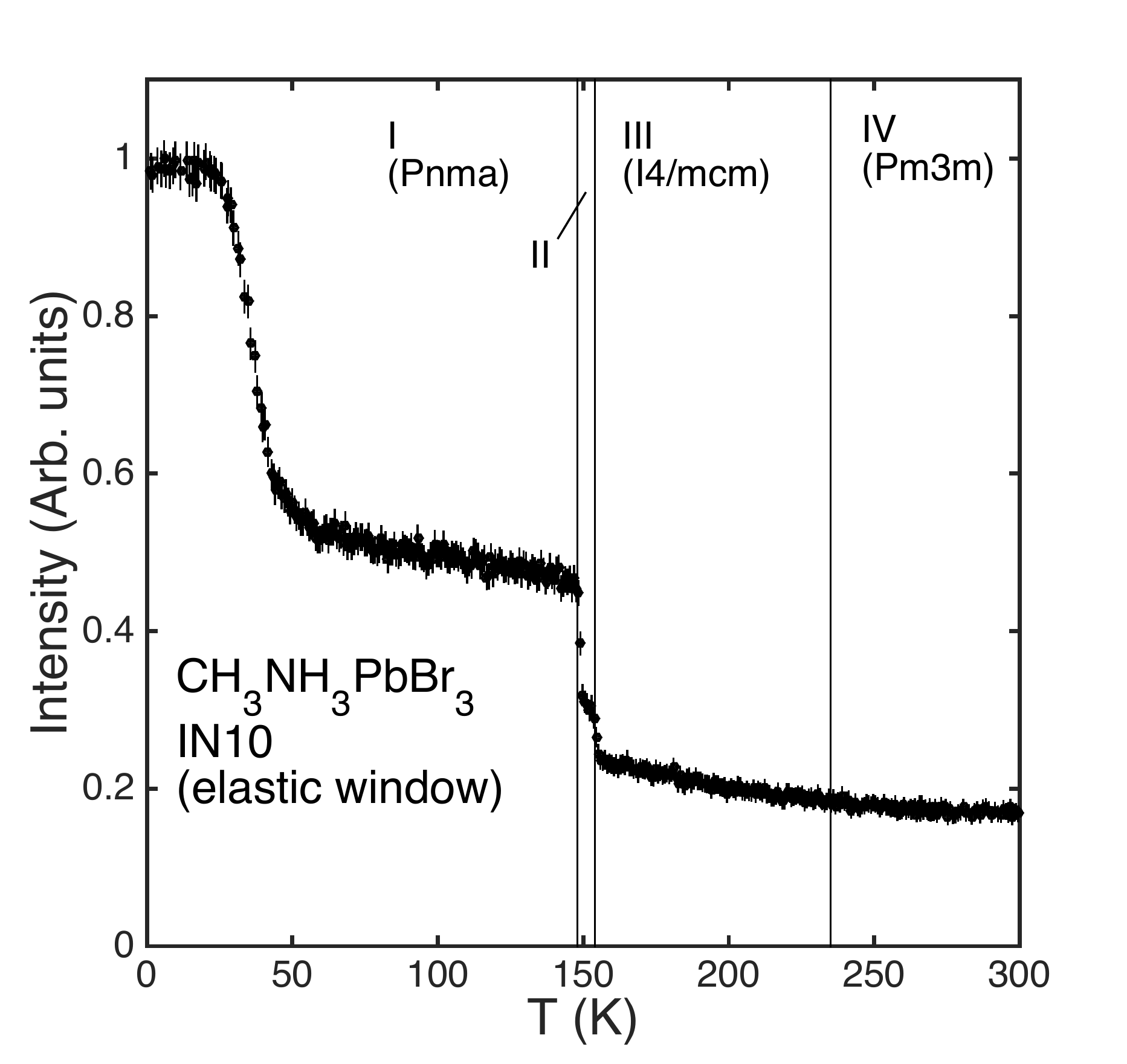}
\caption{\label{window}  A plot of the integrated intensity of a ``fixed window" scan taken on IN10.}
\end{figure}

Given the limited temperature range where dynamics are observable on IN10 (see Figure 2 of the main text), we then extended these measurements using the NEAT chopper spectrometer (HZB, Berlin).  NEAT is a direct geometry cold chopper spectrometer.  Two different configurations were used with incident wavelengths fixed at 6.3 \AA\ and 12 \AA\ with energy resolutions of $\delta E\sim$ 8 $\mu eV$ and 70  $\mu eV$ respectively.  The results of these measurements are combined with the IN10 data in Fig. 2 of the main text.

A combination of both IN10 and NEAT provide a temperature dependence of the dynamics up to the structural transition from an orthorhombic to tetragonal unit cell.  However, the dynamics were found to quickly leave the energy window at the structural transition at 150 K.  To probe the dynamics over a broader energy range with coarser energy resolution we used the IN4 direct geometry thermal Fermi chopper spectrometer (ILL, France).   The incident energy was fixed to 16.84 meV for the data presented.

To finally probe all of the dynamics as a function of temperature and to provide a dataset amenable for comparisons with calculations (to determine the nature of the soft phonons) we performed a series of experiments on the MARI direct geometry chopper spectrometer at ISIS.  For the data presented in Fig. 4 of the main text, we used a Gd Fermi chopper to fix the incident energy in combination with a disc chopper to remove background from high energies.  These experiments used E$_{i}$=25 meV, 50 meV, and 100 meV spinning the fermi chopper at 400 Hz, 400 Hz, and 600 Hz respectively.    To investigate the internal modes presented here in the supplementary information, the Gd fermi chopper was used with E$_{i}$=150 meV spin at 300 Hz.  For higher energies, the relaxed fermi chopper was used with E$_{i}$=250 meV, 550 meV, and 1000 meV.  The chopper was spun at 300 Hz, 300 Hz, and 600 Hz respectively.   For each spectrometer configuration, we used the low-temperature ($T$=2 K) dataset as the resolution function, and convolved this with the model $S(Q,E)$ after background subtraction.

\section{``Elastic Window" Scan}

\renewcommand{\thefigure}{S2}
\begin{figure}[t]
\includegraphics[width=7.5cm] {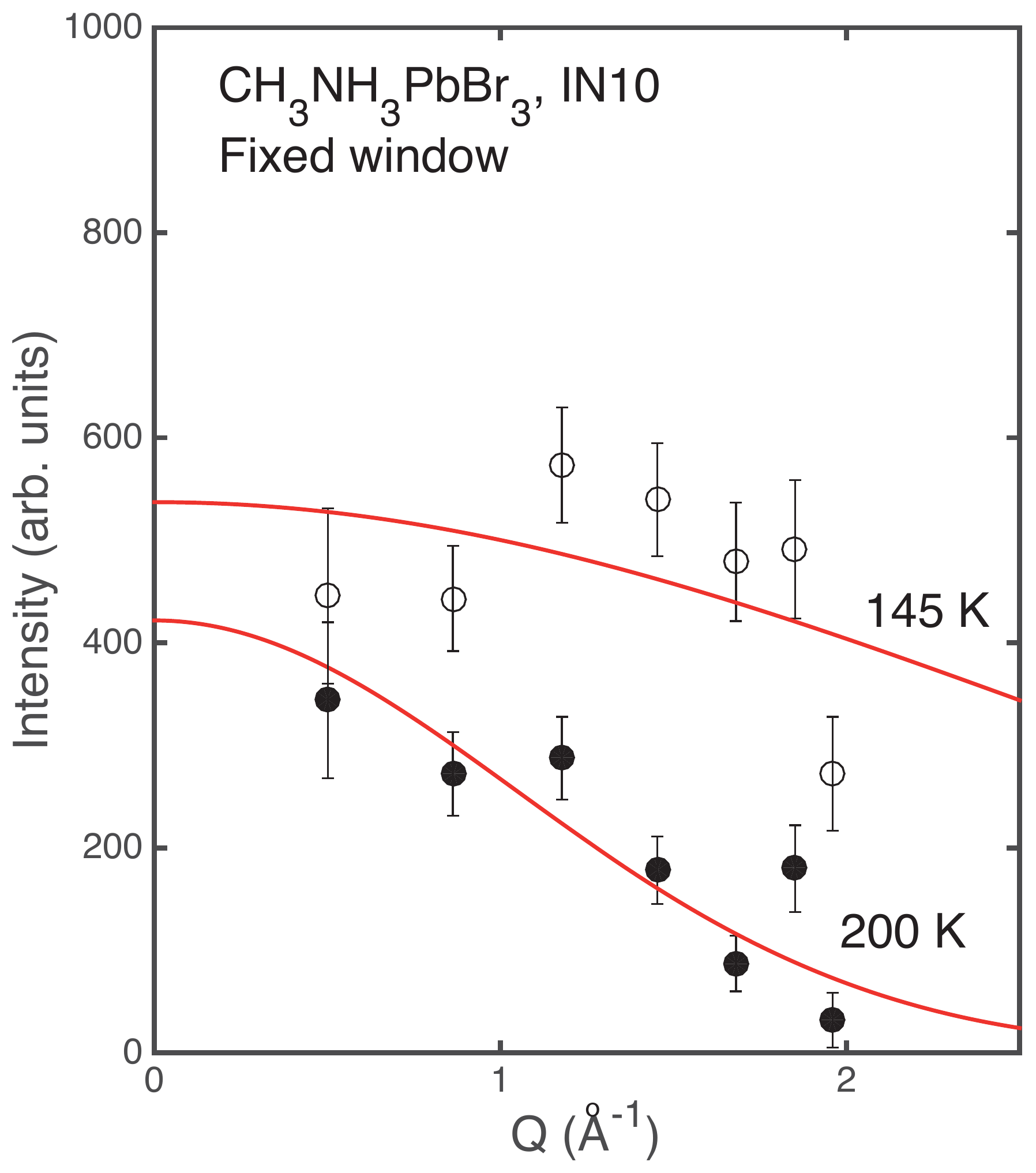}
\caption{\label{fixed_q}  A sample plot of the momentum dependence extracted from a ``fixed window" scan done on the IN10 backscattering spectrometer at 145 K and 200 K.  The solid lines are fits to gaussians used to extract $\langle u^{2}\rangle$ as a function of temperature.  The results of this fit are plotted in Fig. 1 of the main text.}
\end{figure}

In the main text, an elastic window scan from the IN10 backscattering spectrometer was used to estimate $\langle u^{2} \rangle$.  In this scan, the instrument integrates over the elastic line and $\pm$ 10 $\mu eV$.  The integrated intensity summed over all detectors is shown in Fig. \ref{window}.  The intensity can be seen to show discontinuities at the same temperatures illustrated in Fig. 1 in the main text.  The decrease in intensity at $\sim$ 50 K does not correspond to any transition in the bulk material and neutron inelastic scattering work (presented in the main text) finds this to be due to methyl group rotations.

We have used this scan to estimate $\langle u^{2} \rangle $ by fitting the momentum dependence at each temperature to the form $I(Q)\sim e^{-\langle u^{2}\rangle Q^{2}}$.  The IN10 backscattering spectrometer consists of seven detectors covering a momentum range from 0.4 \AA$^{-1}$ to 1.96 \AA$^{-1}$ and example fits of the ``fixed window" scan are shown in Fig. \ref{fixed_q}.  The results of these fits at each temperature are shown in Fig. 1 of the main text.

\section{High-energy internal modes}

We now discuss the higher energy lattice modes probed with MARI.  Figure \ref{mari} shows there to be considerable band gap in the lattice excitations between $\sim$ 40 meV and $\sim$ 110 meV.  This gap separates low-energy modes involving the inorganic framework and the molecular excitations and the higher energy internal modes of the MA molecule.   As shown in the main text, while the lower energy band of excitations is highly temperature dependent, the higher energy modes show comparatively little temperature dependence.  Here we use previous Raman spectroscopy results on CsPbCl$_{6}$~\cite{Carabatos03:34} and the antifluorite (MA)$_2$SnCl$_6$~\cite{Shi98:29} to distinguish the two different sets of vibrations and to corroborate the Castep analysis used in the main text to help separate lattice modes from the PbBr$_{6}$ framework and the molecular motions. 

\renewcommand{\thefigure}{S3}
\begin{figure}[t]
\includegraphics[width=7.0cm] {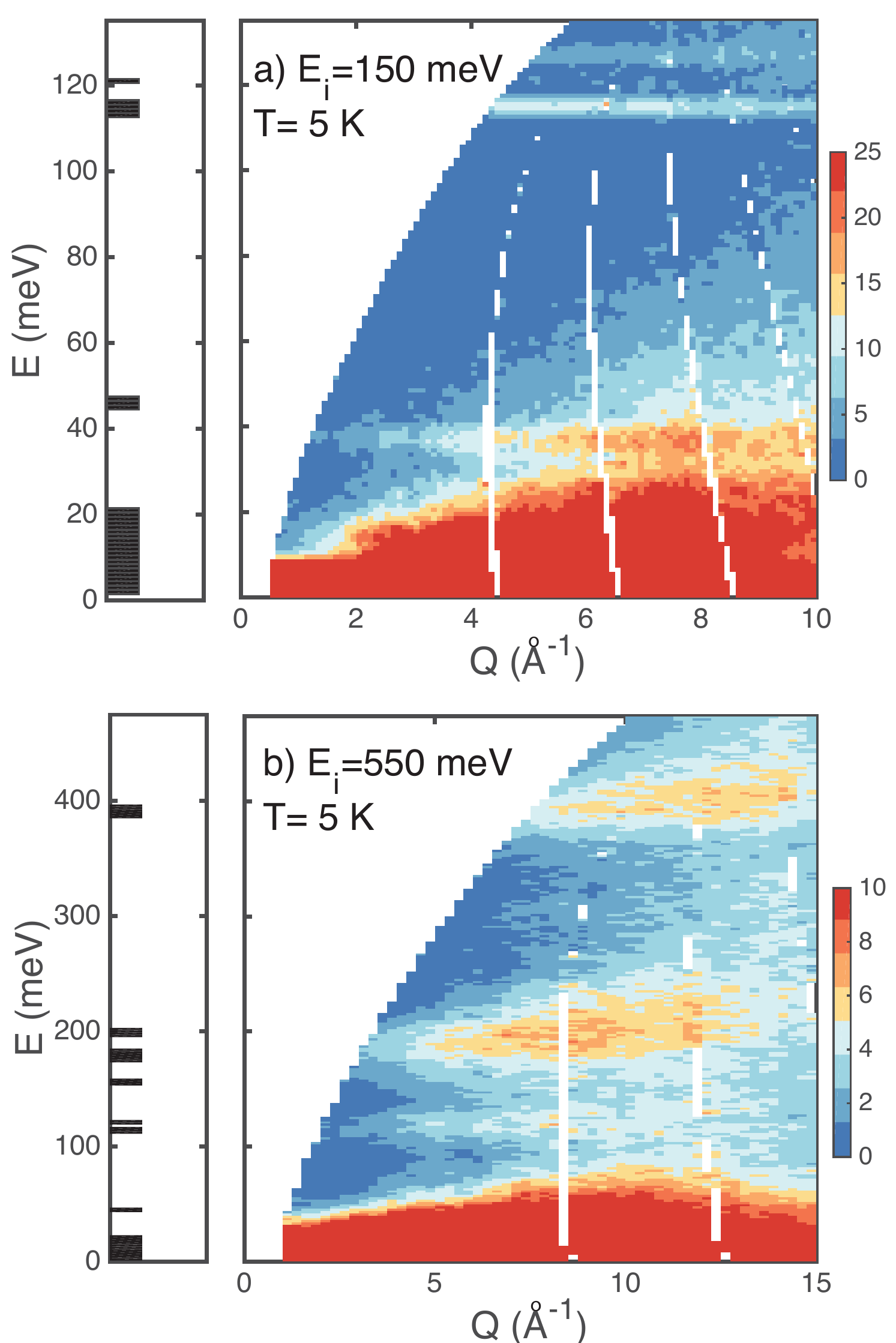}
\caption{\label{mari}  High energy data on MARI illustrating the two bands of excitations from modes involving the PbBr$_{6}$ framework and the internal molecular motions at high energies.  The two panels illustrate data taken with E$_{i}$=150 meV and 550 meV respectively.  The data is compared to Castep calculations on the left illustrating where a finite neutron cross section is expected.}
\end{figure}



Based on previous studies of (MA)$_{2}$SnCl$_{6}$,~\cite{Shi98:29}, we reproduce the mode assignment presented in Table \ref{table}.   The mode energies are comparable to that observed here for CH$_{3}$NH$_{3}$PbBr$_{3}$ and are corroborated by calculations performed with Castep.


\begin{table}[ht]
\caption{Assignment of the high-frequency modes from the 5~K spectra taken on MARI. Frequencies are from Raman studies on (MA)$_{2}$SnCl$_{6}$ at 8 K.~\cite{Shi98:29}}
\centering
\begin{tabular} {c c c c c }
\hline\hline
$\nu$ & Nature & Energy (meV)\\
\hline\hline
$\nu_{7}$    &  N-H antisymmetric strtech 				& 400  			\\
$\nu_{1}$    &  N-H symmetric stretch 				& 397 			\\
$\nu_{8}$    &  C-H antisymmetric stretch  				& 372 			\\
$\nu_{2}$    &  C-H symmetric stretch 				& 360 			\\
$\nu_{9}$    &  NH$_{3}$ antisymmetric deformation 		& 198 			\\
$\nu_{3}$    &  NH$_{3}$ symmetric deformation 		& 182 			\\
$\nu_{10}$  &  CH$_{3}$ antisymmetric deformation 		& 180 			\\
$\nu_{4}$    &  CH$_{3}$ symmetric deformation 		& 176 			\\
$\nu_{11}$  &  rock								& 155 			\\
$\nu_{5}$    &  C-N stretch 						& 121 			\\
$\nu_{12}$  & rock								& 112			\\
\hline
\label{table}
\end{tabular}
\end{table}

The study of modes involving the lead halide perovskite structure have been performed by several groups including Raman studies in Ref. \onlinecite{Calistru97:82,Carabatos03:34}.  These mode assignments are expected to differ from what is observed here owing to the different between chlorine and bromine.  We also note that Raman spectra in CsPbCl$_{6}$ only shows a comparatively small softening of the low-energy modes in comparison to what we report in the main text for CH$_{3}$NH$_{3}$PbBr$_{3}$.  However, the results qualitatively agree with the energy range of the lattice dynamics that we report in the main text.

\textit{Open data access:} Following UK Research Council guidance, data files can be accessed either at source (from ISIS (www.isis.stfc.ac.uk) or the ILL (www.ill.eu)) or through the University of Edinburgh's online digital repository (datashare.is.ed.ac.uk) after publication.



%